\documentclass[a4paper,fleqn,usenatbib]{mnras}

\usepackage{newtxtext,newtxmath}

\usepackage[T1]{fontenc}
\usepackage{ae,aecompl}

\usepackage{graphicx}	
\usepackage{amsmath}	
\usepackage{amssymb}	
\usepackage{cancel}
\usepackage{ulem}
\usepackage{longtable}

\newcommand*{\red}{\textcolor{red}}

\newcommand\Fgamma{\ensuremath{F_\gamma\left(>50\:\mathrm{GeV}\right)}}
\newcommand\Funit{\ensuremath{\mathrm{ph}\,\mathrm{cm}^{-2}\,\mathrm{s}^{-1}}}

\title[Connecting blazars, cosmic rays, and neutrinos]{Connecting blazars with ultra high energy cosmic rays and astrophysical neutrinos}

\author[E. Resconi et al.]{E. Resconi$^{1}$\thanks{E-mail:
elisa.resconi@tum.de}, S. Coenders$^1$, P. Padovani$^{2,3}$, 
 P. Giommi$^{4,5}$, L. Caccianiga$^{6}$\\
$^{1}$Technische Universit{\"a}t M{\"u}nchen, Physik-Department, James-Frank-Str. 1, 
D-85748 Garching bei M{\"u}nchen, Germany\\
$^{2}$European Southern Observatory, Karl-Schwarzschild-Str. 2,
D-85748 Garching bei M\"unchen, Germany\\
$^{3}$Associated to INAF - Osservatorio Astronomico di Roma, via Frascati 33,
I-00040 Monteporzio Catone, Italy\\
$^{4}${Italian Space Agency, ASI, via del Politecnico s.n.c., I-00133 Roma Italy}\\
$^{5}${ICRANet-Rio, CBPF, Rua Dr. Xavier Sigaud 150, 22290-180 Rio de Janeiro, Brazil}\\
$^{6}${INFN sezione di Milano, Dipartimento di Fisica, Universit\'a degli Studi di Milano
via Celoria 16, 20133 Milano, Italy}}

\pubyear{2017}

\begin{document}
\label{firstpage}
\pagerange{\pageref{firstpage}--\pageref{lastpage}}
\maketitle

\begin{abstract}
We present a strong hint of a connection between high energy
$\gamma$-ray emitting blazars, very high energy neutrinos, and ultra high
energy cosmic rays. We first identify potential hadronic sources by filtering
$\gamma$-ray emitters 
in spatial coincidence
with the high energy neutrinos detected by IceCube. The neutrino filtered
$\gamma$-ray emitters are then correlated with  the ultra high energy cosmic
rays from the Pierre Auger Observatory and the Telescope Array by scanning in 
$\gamma$-ray flux ($F_{\gamma}$) and angular separation ($\theta$) between sources and
cosmic rays.  A maximal excess of 80 cosmic rays (42.5 expected) is
found at $\theta\leq10^{\circ}$  from the neutrino filtered $\gamma$-ray
emitters selected from the second  hard {\it Fermi}-LAT catalogue (2FHL)
and for $\Fgamma\geq1.8\times10^{-11}\:\Funit$.  The probability for this to happen
is $2.4 \times 10^{-5}$, which translates to $\sim 2.4 \times 10^{-3}$ after
compensation for all the considered trials.  No excess of cosmic rays is instead observed for  the
complement sample of $\gamma$-ray emitters (i.e. {\it not} in spatial
connection with IceCube neutrinos). A likelihood ratio test comparing the
connection between the neutrino filtered and the complement source samples with the
cosmic rays favours a connection  between neutrino filtered emitters and cosmic
rays with a probability of $\sim1.8\times10^{-3}$ ($2.9\sigma)$ after
compensation for all the considered trials.  The neutrino
filtered $\gamma$-ray sources that make up the cosmic rays excess are blazars
of the high synchrotron peak type. More statistics is needed to further investigate 
these sources as candidate  cosmic ray and neutrino emitters. 
\end{abstract}




\section{Introduction}
\label{sec:Introduction}

Blazars are Active Galactic Nuclei (AGN) hosting a jet oriented at a small
angle with respect to the line of sight with highly relativistic particles
moving in a magnetic field and emitting non-thermal radiation
\citep{Urry:1995mg}. They are among the most luminous and most energetic sources in
the Universe. The spectral energy distributions (SEDs) of blazars are composed
of two broad humps, a low energy and a high energy one. 
Blazars are subdivided in two main
sub-classes, namely flat-spectrum radio quasars (FSRQ) and BL Lacertae objects
(BL Lacs), with the former displaying strong, broad emission lines and the latter instead 
being characterized by optical spectra showing at most weak emission lines, sometimes exhibiting 
absorption features, and in many cases being completely featureless. 
Based on the
rest-frame value of the frequency of the peak of the low energy hump (synchrotron peak, $\nu_S$)
blazars of the BL Lac type are called high energy peaked (HBLs) when $\nu_S > 10^{15}$ Hz
($>4\:\mathrm{eV}$).
HBLs are the rarest types of blazars, making up
$\approx 10\%$ of BL Lacs \citep{Padovani:1994sh}, being at the same time very powerful 
$\gamma$-ray emitters. The idea that blazars of various
types could be sources of ultrahigh energy cosmic rays (CRs) and, related to
that, of high energy neutrinos, has been discussed in
e.g. \cite{Mannheim:1995mm, Halzen:1997hw} and has since then been explored
in a number of studies (see e.g. \citealt{Muecke:2002bi, Kistler:2013my,
Murase:2014foa, Tavecchio:2014eia, Padovani:2015mba, 
Kistler:2016ask}). We perform here detailed
statistical studies to test a blazar connection with CRs going 
through high energy neutrinos used as ``intermediaries'' (see \citealt{1419789}
for an extensive review of how CRs are accelerated up to
$10^{19}-10^{20}$ eV).

\begin{figure}
\includegraphics[width=\columnwidth]{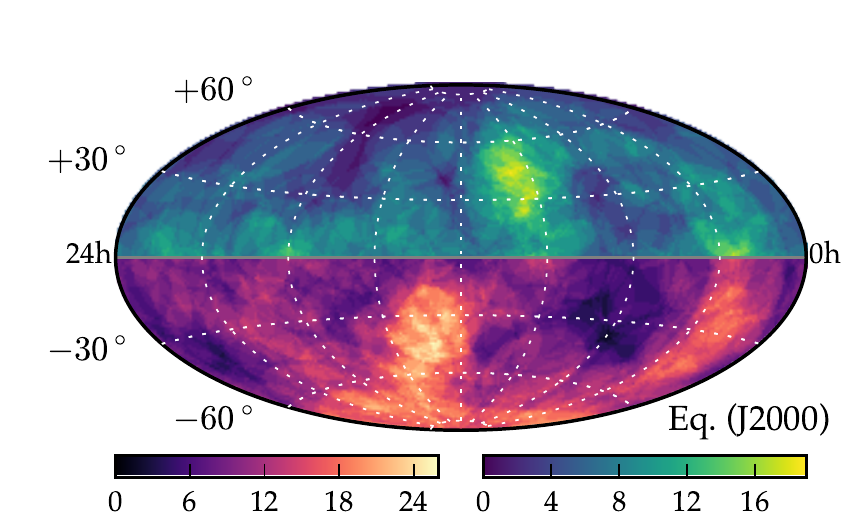}
\includegraphics[width=\columnwidth]{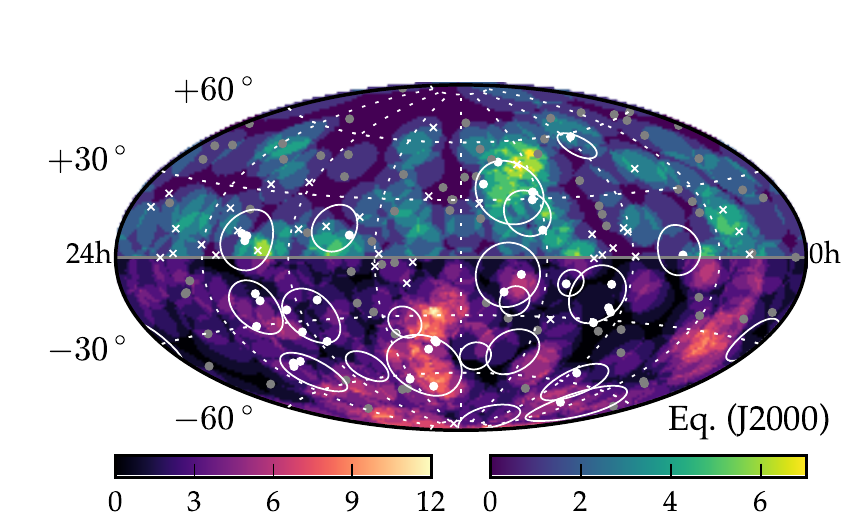}
\caption{Sky maps in equatorial coordinates: (top) Auger and TA data
with uncertainty areas of 20$^\circ$ as reported by the collaborations
(and roughly the size of the two hot spots).  The color scale indicates the
number of events overlapping within the 20$^\circ$ areas; (bottom) Auger and TA
data with uncertainty areas of 10$^\circ$ as obtained from our test with
overlaid neutrinos contributing to the test (crosses are track-like events,
circles are shower-like events with associated median angular error).
Dots indicate the 2FHL (HBL + unclassified sample) objects that are
filtered (white) or not filtered (grey) by neutrinos as described in the text.}
\label{fig:SkyMap}
\end{figure}

The Pierre Auger Observatory  \citep{Abraham:2004dt,
ThePierreAuger:2015rma} and the Telescope Array (TA) \citep{AbuZayyad:2012kk}
have collected together more than 300 CRs
with $E\geq 52 \times 10^{18}\:\mathrm{eV}$ over the entire sky. The former
has detected 231 such events
covering mostly the southern hemisphere and the latter 72 events above
57$\times10^{18}\:\mathrm{eV}$ covering mostly the northern sky. Despite a
series of studies on the CR arrival direction distributions
\citep{PierreAuger:2014yba, Abu-Zayyad:2013vza}, no counterparts have yet been
determined. The strongest deviations from isotropy (post-trial probabilities $\sim 1.3 - 1.4 \%$)
have been obtained for $E \geq 58 \times 10^{18}\:\mathrm{eV}$ in connection with 
Swift AGN closer than 
130 Mpc and more luminous than $10^{44}$ erg s$^{-1}$, and around the direction of Cen A 
(at angular separations
$\sim 18^{\circ}$ and $\sim 15^{\circ}$ respectively) \citep{PierreAuger:2014yba}. 
The distribution of CRs does not present statistically
significant small scale structures. 
However, two ``hot'' regions ($2-3\sigma$
deviations) of about $20^\circ$ in the CR sky have been reported 
(see Fig.~\ref{fig:SkyMap} top, 
\citealt{Abbasi:2014lda}, and \citealt{PierreAuger:2014yba}) but their interpretation is not yet
clear. 

The IceCube South Pole Neutrino Observatory has recently discovered high energy
astrophysical neutrinos and has reported a sample of 54 events
collected over a period of four years with a deposited energy up to
$2\:\mathrm{PeV}$~\citep{Aartsen:2013jdh, Aartsen:2014gkd,
2015arXiv151005222T}. These events are coming from the entire sky and 
make up what we define as the Isotropic Neutrino Emission (I$\nu$E). They
consist of neutrinos of all flavors, which interact inside the instrumented
volume 
(starting events),
the majority of which are shower-like\footnote{Shower-like events are
one of the topologies for neutrino observatories. Showers have a good energy
resolution but reduced angular resolution ($\sim10^\circ$). The other
topology are track-like events induced by muons, that have much better angular
resolution ($<1^\circ$), in which not all the energy is deposited on the detector.}.
The complementary sample of
through-going charged current 
$\nu_\mu$ from the northern sky has also been 
studied over a period of six years 
and recently been reported in \cite{Aartsen:2016xlq} showing that their
spectrum is inconsistent with the hypothesis of purely terrestrial origin at the 
$5.6\sigma$ level. 
The track-like events confirm the general picture of the
I$\nu$E although their energy spectrum E$^{-\gamma}$ is harder ($\gamma = 
2.13 \pm 0.13$) with respect to the all sky one obtained from the starting event
sample ($\gamma = 2.58 \pm 0.25$), suggesting a mixed origin of the signal
observed by IceCube. Many diverse hypothesis for the astrophysical
counterparts of the I$\nu$E have been put forward (see, e.g. 
\citealt{2016MNRAS.455..838K, 2016arXiv160203694M, 
2016PhRvL.116g1101M, 2016PhRvD..93e3010T, 2016PhRvD..93h3003S, 2016PhRvD..93h3005W,
2016APh....78...28Z, 2016PhRvD..93l3002N}) but none has so far been
statistically supported by the observational data.

In \cite{Padovani:2016wwn} the authors have correlated the second catalogue of
hard {\it Fermi}-LAT sources (2FHL) ($E>50\:\mathrm{GeV}$, 360 sources of
various types, mainly blazars, \citealt{Ackermann:2015uya}), and two other 
catalogs (see Appendix), with the publicly available high energy neutrino sample detected by
IceCube (\citealt{Aartsen:2014gkd, 2015arXiv151005222T, Aartsen:2015rwa,
2015ATel.7856....1S}, including only events with energy and median angular error $\ge 60$ TeV 
and  $\le 20^{\circ}$ respectively). The probability scanning over \Fgamma~was 0.4\%.
We have now also evaluated the impact of the trials, which results in a global
p-value of 1.4\% ($2.2\sigma$). This applies only to HBL blazars and appears
to be strongly dependent on $\gamma$-ray flux.

In \cite{Aartsen:2015dml}, the three collaborations IceCube, Auger, and TA 
jointly reported about correlation tests between very high energy neutrinos and
the same sample of CRs used in this paper. All the studies were compatible with the
null hypothesis of no correlation although two interesting excesses were
observed when considering IceCube shower-like events: a post-trial probability of
0.05\% for an angular scale of 22$^\circ$ provided by the comparison to a 
randomized CR sky, and a probability of 0.85\% 
obtained by
scrambling the neutrinos, thus preserving the existing anisotropies observed
in the arrival directions of the CRs.
As also discussed in  \cite{Aartsen:2015dml}, the first result (CR randomization),  is affected by the presence
of anisotropies (hot spots) in the CR sky (see Fig.~\ref{fig:SkyMap}, bottom). 
More interesting is the second result (neutrino randomization), which hints at a possible connection between IceCube neutrinos and CRs.

\begin{figure}
\includegraphics[scale=1.0]{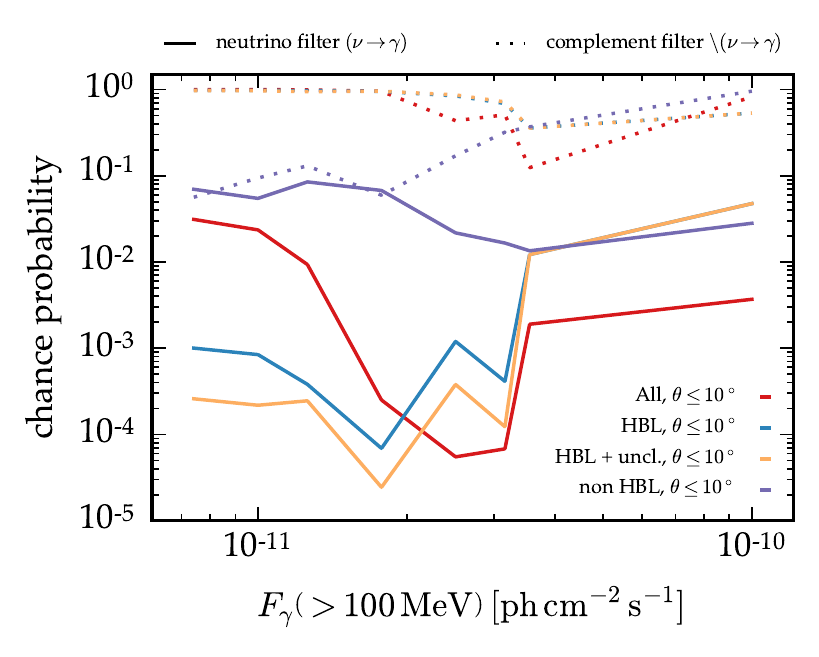}
\includegraphics[scale=1.0]{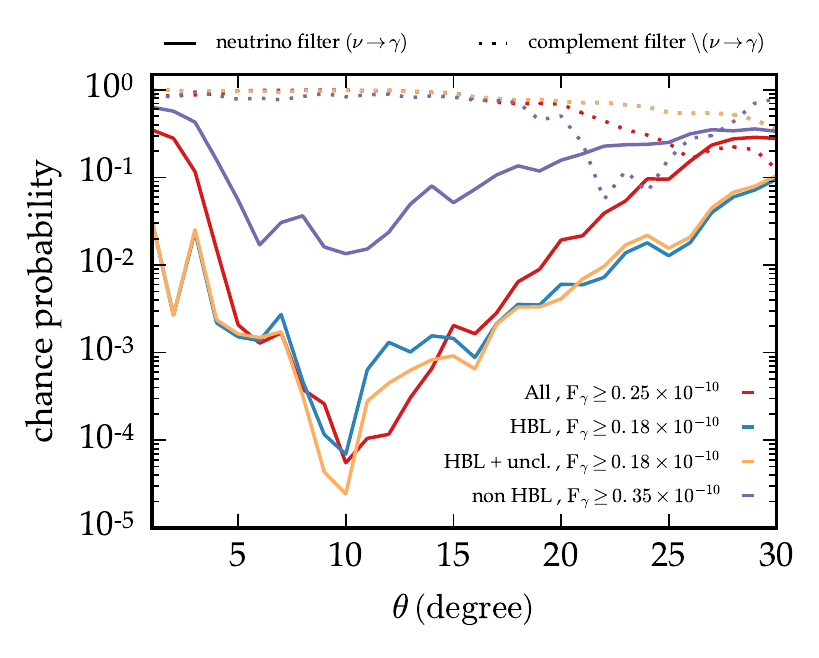}
\caption{($\nu \rightarrow \gamma  \rightarrow$ CR) correlation test results (2FHL): (top) $F_\gamma$  projection of the  chance
probability at a fixed angular
distance~$\theta$. 
(bottom) $\theta$  projection of the chance
probability at a fixed $F_\gamma$ value.
The probabilities are reported for the neutrino filtered $\gamma$-ray emitters (solid line) and for the complement sample (dashed line) at $F_\gamma$ (or $\theta$) larger
than the value on the x-axis.}
\label{fig:Pvalues}
\end{figure}

\section{Statistical Analysis}
\label{sec:analysis}
Motivated by the hints mentioned above on (1) an HBL origin of some of the
IceCube neutrinos from \cite{Padovani:2016wwn}, and (2) a common origin of neutrinos
and CRs from \cite{Aartsen:2015dml}, we have developed a two-step analysis to
investigate the connection between $\gamma$-ray emitters and CRs. 

\subsection{Neutrino filter to \boldmath$\gamma$-ray sources (\boldmath $\nu \rightarrow$ $\gamma$) }
\label{step1}
The neutrinos are implied here as {\it filters}
in order to 
single out the best candidate lepto-hadronic accelerators
\citep{Petropoulou:2015upa} from $\gamma$-ray catalogs and hence the most probable
CR sources. We have used all the high energy catalogs presently available
including the 2FHL, the 2WHSP~\citep{Chang:2016mqv}, and the 3LAC~\citep{Ackermann:2015yfk}, as detailed in
\cite{Padovani:2016wwn} and Tab.~\ref{tab:results}, focusing on HBL
blazars\footnote{The determination of $\nu_S$, which is required to classify
sources as HBLs, requires the availability of multi-frequency data and is also
affected by variability, a defining feature of blazars. The HBL - non-HBL
distinction is therefore not sharp by definition.}.
We have also considered the 2FHL and 3LAC catalogue subsets as reported. 
The neutrino list is
composed of the 51 events (30 starting and 8 tracks) selected by
\cite{Padovani:2016wwn} and the recently published 29 through going tracks
reported in \cite{Aartsen:2016xlq} of which one is a starting track
also. No further catalogue has been tested and all the tests performed on the
data are reported here not to hide any trial or relevant information.

As done in \cite{Padovani:2016wwn}, we have
 filtered $\gamma$-ray sources  in spatial coincidence with the neutrinos (i.e.
within the median angular errors). The selection is done partitioning the flux
of the cataloged sources ($F_\gamma$) (or alternatively the ``figure of merit''
(FoM) introduced in the 2WHSP)\footnote{The FoM is defined as the ratio
between the synchrotron peak flux of a source and that of the faintest blazar
in the WHSP sample already detected in the TeV band.}. Two subsets of the  $\gamma$-ray  catalogues are
then obtained, the neutrino filtered and its complement:
\begin{align}
    \qquad&(\nu \rightarrow \gamma);F_{\gamma} \hspace{0,2cm}neutrino \hspace{0,1cm}filter,\\
    \backslash&(\nu \rightarrow \gamma);F_{\gamma} \hspace{0,2cm}complement.
\label{eq:nu_filter}
\end{align}

With respect to
\cite{Padovani:2016wwn}, we have here extended the coverage of the sky for 2FHL
to the Galactic plane to allow the realization of neutrino scrambled maps as background cases and included the recently published through going tracks.

\subsection{Neutrino filtered sources to CRs\\ \hspace{0.7cm} (\boldmath$\nu \rightarrow \gamma  \rightarrow$ CR)}
\label{step2}
In a second independent step, the neutrino selected $\gamma$-ray sources at various
$F_\gamma$ (or FoM) are cross correlated with the CRs. The correlation is also 
done as a function of the angular separation $\theta$ between the source and
the reconstructed incoming direction of the CRs over the $1^{\circ} - 30^{\circ}$ range
in steps of $1 ^{\circ}$. This is because the CRs are charged particles and therefore deflected by 
an unknown angle due to the intervening magnetic fields.

We quantify the strength of the correlation by counting the number of CR
events that have {\it at least} one neutrino  filtered $\gamma$-ray source with 
$F_\gamma$ at distance smaller than the angular distance~$\theta$ and compare
this to random trials. This is done by randomizing the right ascension of the 
neutrino events and repeating the statistical analysis mentioned above. This
randomization method is regularly used within the IceCube collaboration and ensures that the aforementioned anisotropies in the CRs are
conserved and do not artificially contribute to the significance of the result, 
preserving at the same time the IceCube neutrino distribution, which is known to be 
not uniformly distributed. 


The final probabilities  are calculated based on random trials. The final result is corrected for
trials due to the scanning in $F_\gamma$ (or FoM) and $\theta$ to search for the largest
CR excess over the random expectation. We have also corrected for the number of subsets in the 2FHL considering the relative correlations within the subsets. 
The additional factor is 2.9.  

Although the CR horizon is limited by the energy losses caused by the
interactions of CRs with photons of the Cosmic Microwave Background (CMB),
 the so-called GZK (Greisen-Zatsepin-Kuzmin) effect
(\citealt{Greisen:1966jv, Zatsepin:1966jv}), we do not apply a-priori cuts on the
distance of the cataloged sources. The reason is that a large number of BL Lacs
has no measured redshift due to the lack of emission lines in their optical
spectra (see also Sec.~\ref{discussion}).
Moreover, it is difficult to quote a CR horizon a priori.
Instead of scanning in redshift $z$ as done in other tests as the one in \cite{PierreAuger:2014yba} and in  
\cite{Abu-Zayyad:2013vza}, we
use all sources regardless of their redshift.

\subsection{Likelihood ratio test}
\label{LLH}
As a last step, a likelihood ratio test comparing the connection between the filtered ($\nu \rightarrow \gamma$) and the complement $\backslash (\nu \rightarrow \gamma$) source samples with the CRs is performed. We define the test statistics as
\begin{equation}
\hspace{2,0cm}\Lambda=\frac{\mathcal{P}\left((\nu \rightarrow \gamma);F_{\gamma} \rightarrow CR;\theta\right)}
 {\mathcal{P}\left(\backslash (\nu \rightarrow \gamma);F_{\gamma} \rightarrow CR;\theta\right)}, 
\label{eq:llh_ratio}
\end{equation}
where $\mathcal{P}$ is the probability obtained in the two-step statistical test described above. Large values
of $\Lambda$ indicate preference for a stronger correlation of neutrino filtered $\gamma$-ray emitters and 
CRs supporting a physical connection among the three astronomical messengers. We estimate the significance of this test by comparing
data to trials with randomized neutrinos as discussed in the previous section.
The p-value is
then defined as the chance probability that trials produce a test statistics
$\Lambda$ that is larger than, or equal to that observed.


\begin{table*}
\caption{($\nu \rightarrow \gamma  \rightarrow$ CR) correlation test results: local and
global p-values obtained with respect to the $\gamma$-ray catalogs tested. The
local p-value is the minimum p-value observed partitioning the $\gamma$-ray
catalogues in $F_\gamma$ (or FoM) and scanning in angular distance $\theta$ between
neutrino spatially selected sources and CRs. The global p-value is the
corresponding one penalized for the relative trials applied through the two
dimensional scan. The p-value calculation is done using $10^6 - 10^7$
trials depending on the significance of the result. The results of the
likelihood ratio test in Eq.~\eqref{eq:llh_ratio} and the p-value of the
outcome are listed in the last two columns.}
\begin{tabular}{l r | c c c c c c c c c}
$\gamma$-ray & \# & $F_\gamma$   & $\#(\nu\rightarrow\gamma);F_{\gamma}$ & $\theta$ & \multicolumn{2}{c}{\#CRs}                 & local p-value & global p-value                                                 & $\log_{10}\Lambda$ & $\Lambda$\\
catalogue      &    & (FoM) & (FoM)           & (deg)    & observed                  & expected      & \multicolumn{2}{c}{$((\nu\rightarrow\gamma);F_{\gamma}\rightarrow CR;\theta$)} &                    & p-value\\
\hline\rule{0pt}{2.5ex}
2FHL           & 360  & 0.25 & 46  & 10 & 83  & 46.4  & $5.5\times10^{-5}$ & $2.2\times10^{-3}$ & $3.35$  & $3.1\times10^{-3}$ \\
\:\:\:HBL      & 173  & 0.18 & 34  & 10 & 75  & 40.6  & $6.9\times10^{-5}$ & $2.0\times10^{-3}$ & $3.77$  & $1.6\times10^{-3}$ \\
\:\:\: + uncl. & 186  & 0.18 & 36  & 10 & 80  & 42.5  & $2.4\times10^{-5}$ & $8.4\times10^{-4}$ & $4.17$  & $(6.3\times10^{-4})^* $ \\
\:\:\:non-HBL  & 174  & 0.35 & 20  & 10 & 37  & 20.3  & $1.4\times10^{-2}$ & $1.7\times10^{-1}$ & $0.62$  & $2.5\times10^{-1}$ \\
2WHSP          & 1681 & 2.51 & 11  & 17 & 75  & 31.2  & $3.0\times10^{-4}$ & $1.1\times10^{-2}$ & $2.57$  & $1.2\times10^{-2}$ \\
3LAC           & 1444 & 0.56 & 172 & 18 & 231 & 203.4 & $3.3\times10^{-2}$ & $3.3\times10^{-1}$ & $0.42$  & $4.2\times10^{-1}$ \\
\:\:\:HBL      & 386  & 0.71 & 25  & 16 & 131 & 93.4  & $5.6\times10^{-3}$ & $1.1\times10^{-1}$ & $1.53$  & $8.2\times10^{-2}$ \\
\:\:\:FSRQ     & 415  & 1.00 & 50  & 18 & 174 & 166.6 & $3.6\times10^{-1}$ & $9.2\times10^{-1}$ & $-1.11$ & $8.6\times10^{-1}$ \\
\:\:\:other    & 645  & 1.00 & 48  & 3  & 19  & 12.1  & $4.0\times10^{-2}$ & $3.6\times10^{-1}$ & $0.08$  & $4.4\times10^{-1}$ \\
\hline
\end{tabular}
\\$^*$ We apply a further trial factor of 2.9 to account for the 4 nested sub-catalogues.
\label{tab:results}
\end{table*}

    
\section{Results}
The {\it neutrino filter} $(\nu \rightarrow \gamma)$  introduced in  \cite{Padovani:2016wwn} selects the $\gamma$-ray emitters  which are within the median angular errors of the IceCube neutrino events.
A significant excess of CRs is here reported in connection with the neutrino filtered 2FHL $\gamma$-ray emitters (see Fig.~\ref{fig:Pvalues} solid lines and
Tab.~\ref{tab:results}). The $(F_\gamma, \theta)$ scan provides as the most significant excess a total of 
83 CRs at an angular separation $\theta < 10^\circ$ from the 46 2FHL  neutrino filtered $\gamma$-ray emitters
 with $\Fgamma\geq 2.5\times10^{-11}\:\Funit$. 
This number of CRs has to be compared with an expectation of 46.4 random associations determined on neutrino randomized cases. 
The probability of observing such an excess on randomized maps is 5.5$\times 10^{-5}$ which
translates into 2.2$\times 10^{-3}$ after compensation for trials. 

Among the subsets of $\gamma$-ray emitters reported for the 2FHL, 
the best/minimal probability is obtained for the HBLs +
unclassified sample.  This sample includes all HBLs plus all 
sources with $2^{\circ} < |b_{\rm II}| < 10^{\circ}$, which are still unclassified in the catalogue\footnote{By excluding the Galactic plane, in fact, there is a high chance that these sources are
unrecognized blazars lying in a region of the sky where the density of stellar sources is
still quite high. Despite the fact that the optical counterparts of these $\gamma$-ray sources
are still unknown, upon inspection, and after the test was carried out, it turned out that most 
of them have SEDs of the HBL type.}.
The scan values where the maximal excess is found are $\Fgamma\geq1.8\times10^{-11}\:\Funit$ and 
$\theta = 10^\circ$. At these scan values, 
a total of 80 CRs are associated with the neutrino filtered sample, 
to be compared with an expectation of 42.5 random associations. The probability of
observing such an excess on randomized maps is 2.4$\times 10^{-5}$ which
translates into $8.4\times10^{-4}$ (3.14$\sigma$) after compensation for trials due to the $(F_\gamma, \theta)$
scan. Moreover,
considering the fact that we have tested 4 nested subcatalogues of the 2FHL, the final p-value
is $2.4\times10^{-3}$ (2.8$\sigma$).

\begin{figure}
\includegraphics[scale=1.0]{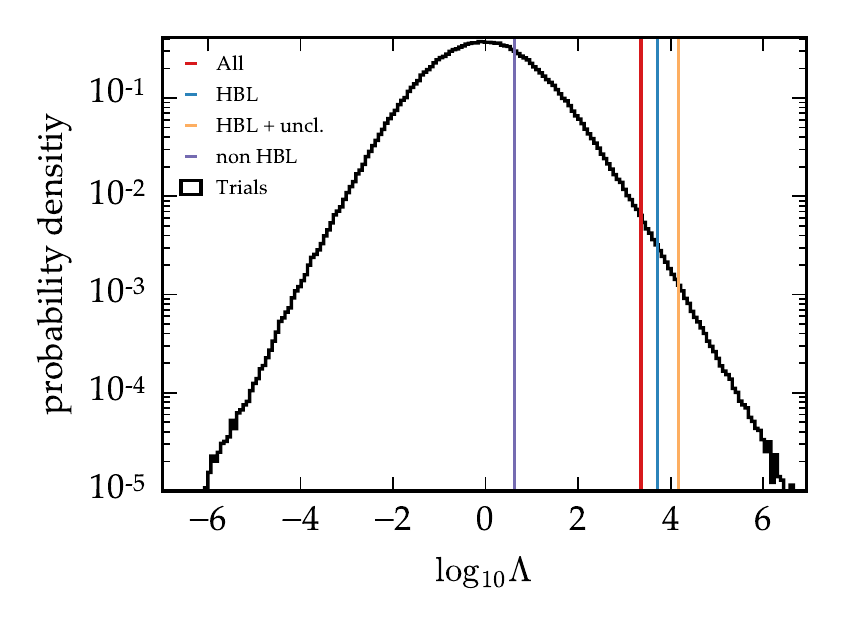}
\caption{\label{LLH ratio} Result of the likelihood ratio test $\Lambda$ (Eq.~\ref{eq:llh_ratio}) for the 2FHL catalogue. Large values of
$\Lambda$ indicate preference of a connection between $\gamma$-rays and CRs for the neutrino filtered $\gamma$-ray emitters. Vertical lines show the outcome of the statistical test for the subsets of the 2FHL catalogue with respect to random trials (black histogram).}
\end{figure}

No excess of CRs is found once the complement sample  of $\gamma$-ray emitters,
$\backslash(\nu \rightarrow \gamma)$, is considered (see Fig.~\ref{fig:Pvalues}
dotted lines).  A likelihood ratio test $\Lambda$ (Eq.~\ref{eq:llh_ratio})
comparing the connection between the filtered and the complement source samples
with the CRs favours a connection  between neutrino filtered emitters  and
CRs (see Fig. \ref{LLH ratio}).  With a p-value of
$1.8\times10^{-4}$ ($2.91\sigma$), the neutrino filter to CRs model is
favoured providing a first strong hint of an association between HBLs and CRs.
More statistics is required to confirm or disprove this scenario.

For the 3LAC and 2WHSP catalogues, we observe similar but less significant excesses
and as for the 2FHL only for HBLs (see
Fig.~\ref{fig:2whsp_Pvalues}, Fig.~\ref{fig:3lac_Pvalues}, Fig.~\ref{fig:2whsp_3lac_logl}
and Tab.~\ref{tab:counterparts} in the Appendix). Also in these cases, no CR excess is observed for the complement sample of neutrino filtered sources (see 
dotted lines in Fig.~\ref{fig:2whsp_Pvalues}, Fig.~\ref{fig:3lac_Pvalues}). The result of the likelihood ratio test described above is
reported in Fig.~\ref{fig:2whsp_3lac_logl} and Tab.~\ref{tab:results}.

\section{Discussion}
\label{discussion}
Out of the 66 IceCube events,  17 are associated with at least one $\gamma$-ray 
counterpart for our best p-value (2FHL HBLs + unclassified sample). 
This is one more than in \cite{Padovani:2016wwn} because here we do not
exclude the Galactic plane; none of these is a track. 
The fact that no neutrino filtered source appears to be the counterpart of a track-like event 
is still compatible with the scenario presented here at the 5\% level.
Hence, if HBLs are neutrinos counterparts they leave room for additional components.
All but one of the neutrino filtered $\gamma$-ray sources have between one and seven CRs associated within
$10^{\circ}$. \\
As mentioned above, we see 80 CRs for an 
expectation of 42.5.  If these numbers are compared to the total of 303 cosmic
rays used in this study, $12.5^{+2.6}_{-2.2}$\% of the entire CR flux
could be interpreted as physically connected to extreme HBLs. This
could be an effect of different magnetic field deflections due to different composition population, 
or additional components in the sources, or a combination of both. 
We note that this value is 
intriguingly close to the
fraction of the IceCube signal explained by the same type of sources ($\sim 10
- 20$\%) from \cite{Padovani:2015mba} and \cite{Padovani:2016wwn}.\\
\indent
Among the selected sources, there are also relatively close-by, well-known, TeV emitters  such as MKN~421 
(also discussed in  \citealt{Fang:2014uja}), PKS~2005--489, and 1ES 1011+496. 
\begin{figure}
    \includegraphics[scale=0.8]{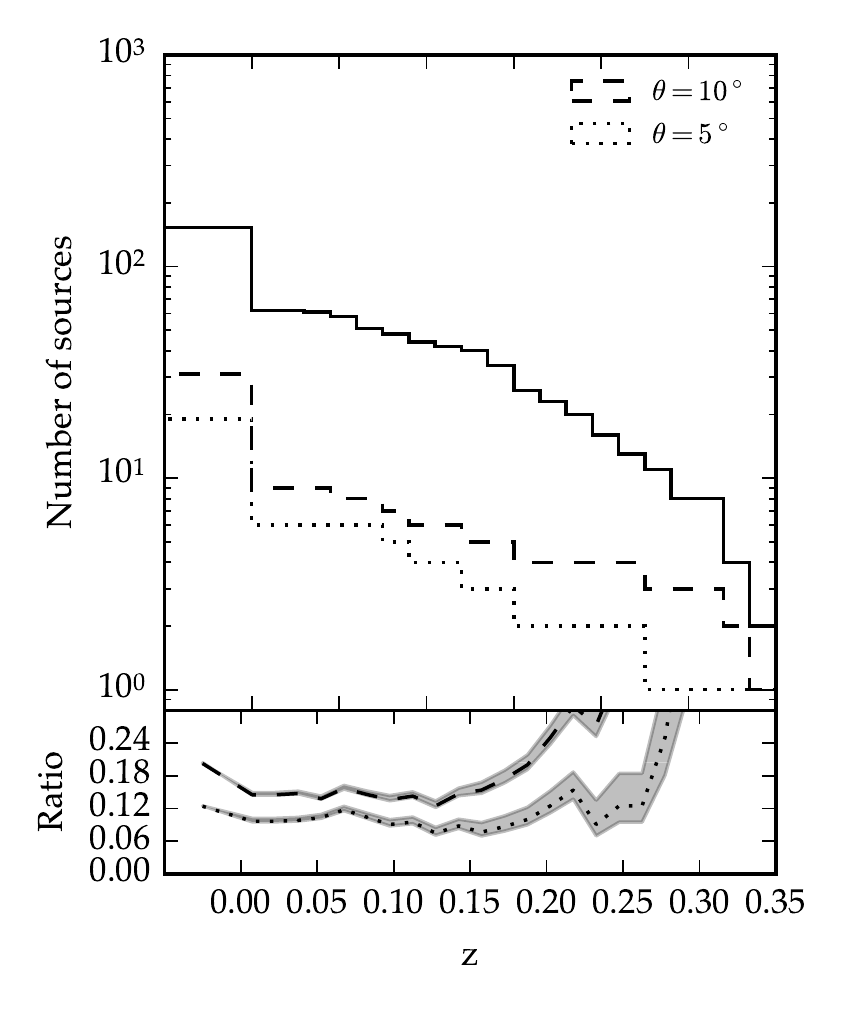}
    \caption{\label{fig:2FHL_z} Top: redshift ($z$) distribution for the  2FHL HBL and
         unclassified objects (continuous line),  at an angular distance $<10^{\circ}$  from the CRs (dashed line), and at  an angular distance $< 5^{\circ}$ from the CRs (dotted line). 
         The sources without a measured $z$  are counted in the bin below zero. Bottom: ratio between the two lower lines  and the upper line in the top plot. }
\end{figure}
Given the redshifts of some of the sources (Fig.~\ref{fig:2FHL_z})
the scenario proposed here 
might seem to be in some tension with the expected GZK suppression (and other CR-background 
interactions). However:
1. the majority of the HBLs do not have a distance determination (and therefore their redshift distribution 
is not very informative); 2.  about half of the sources are expected to be spurious; 3. the intrinsic source power in CRs is unknown. \\
\indent
CRs propagating through space will contribute to the
diffuse $\gamma$-ray emission via pion photoproduction on the CMB photons.
Hence, the scenario proposed here 
has to be questioned also on these terms. Recent work
\citep{Gavish:2016tfl} has discussed the contribution of HBLs to the Extragalactic Gamma Ray
Background (EGRB) detected by {\it Fermi}. 
Given that the EGRB is dominated by
HBLs \citep{Giommi:2015ela, Ajello:2015mfa}, there is little room to
accommodate the inevitable diffuse contribution from $\gamma$-rays resulting 
from CR losses during
propagation. 
Once blazar evolution is correctly
taken into account as done in \cite{Gavish:2016tfl}, however, the secondary $\gamma$-ray 
radiation expected from HBL blazars, which do not show cosmological evolution and 
are a small subset of the entire blazar population,   
is much lower than for other classes leaving these objects as
viable sources of CRs. At the same time, the expected cosmogenic (GZK) 
neutrino flux would also be very low \citep{Taylor:2015rla}.\\
\indent

\section{Conclusion} 
We have provided a first hint (p-value $= 0.18\%$; 2.91$\sigma$) that a
sub-class of blazars may be the counterpart of some high energy neutrinos and
CRs.  These are extreme blazars, i.e. strong, very high energy $\gamma$-ray
sources of the high energy peaked type. Due to the limited angular resolution
of the IceCube neutrinos and the intrinsic magnetic bending of the CRs, one
cannot determine at present which of the counterparts selected is indeed
responsible for the multi-messenger emission. Larger samples of CRs, including
new data and events with $E<52\times10^{18}\:\mathrm{eV}$, and of neutrinos,
would allow us to test further and with better statistics this first strong
hint but unfortunately these data are still proprietary. We therefore suggest
the international collaborations to follow up on this or alternatively make
their catalogs publicly available as they should be.

\section*{acknowledgments}
We thank the many teams, which have produced the data
and catalogues used in this paper for making this work possible. ER is supported
by the Heisenberg Program of the Deutsche Forschungsgemeinschaft (DFG RE
2262/4-1, SFB1258), SC by the cluster of excellence ``Origin and Structure of the
Universe'' of the Deutsche Forschungsgemeinschaft.


\appendix

\section{Results for the 2WHSP and 3LAC catalogues.} 
Table~\ref{tab:results} lists the
results for all catalogues used in this work. As highlighted in the discussion
section, a strong correlation for HBL type objects in the 2FHL catalogue is
found. The 2WHSP catalogue~\citep{Chang:2016mqv} consists only of blazars of the HBL type
excluding the Galactic Plane ($\left|b_\mathrm{II}\right|\geq10^\circ$). The
best (lowest) p-value is found for a ``figure of merit'' $\mathrm{FoM}\geq2.51$
and CR angular distance~$\theta = 17^\circ$. 75 CR events are observed
above an expectation of 31.2. 
\begin{figure}
    \includegraphics[scale=1.0]{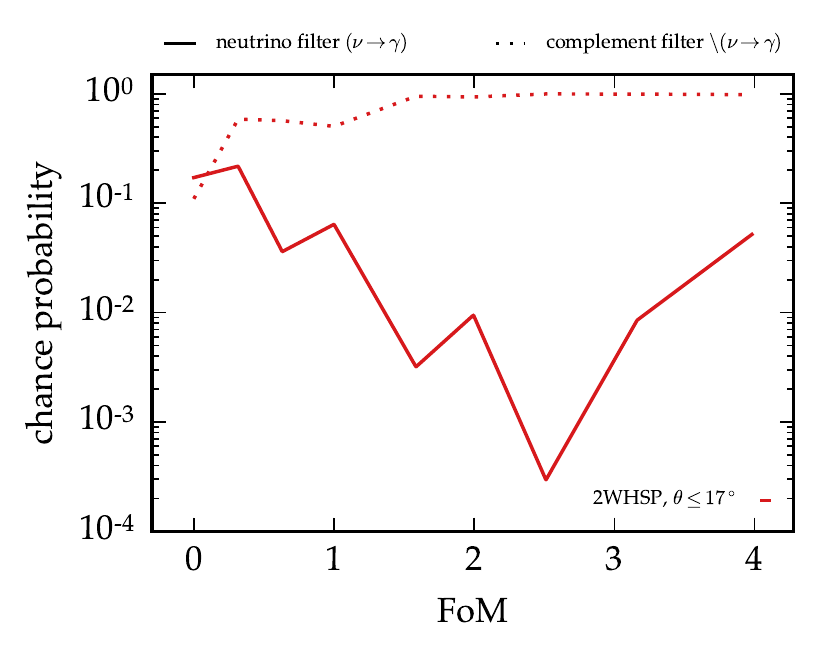}
    \includegraphics[scale=1.0]{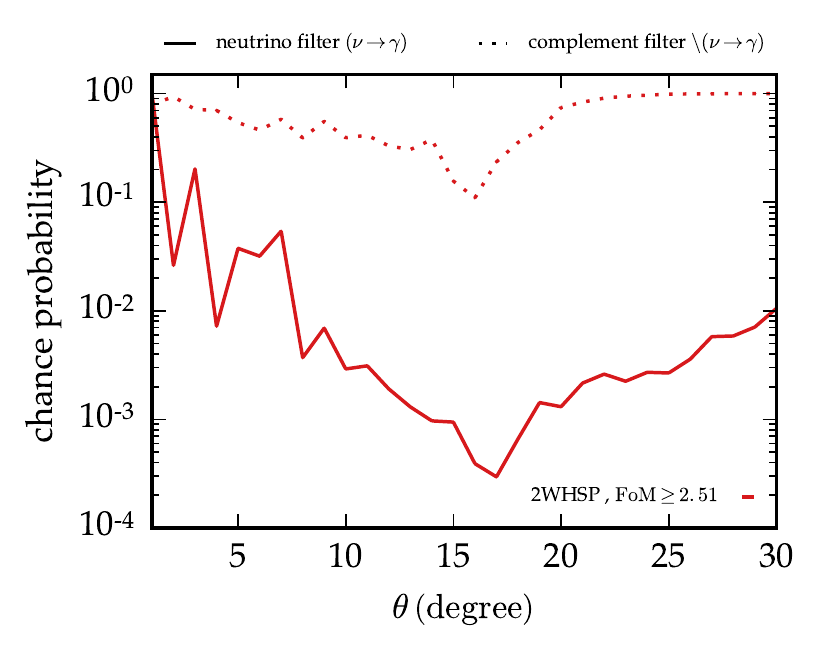}
    \caption{\label{fig:2whsp_Pvalues} Result of the statistical test for the
             2WHSP catalogue scanning over ``figure of merit'' (FoM, top) and 
             CR angular distance $\theta$ (bottom). The probabilities are reported for the neutrino 
             filtered $\gamma$-ray emitters (solid line) and for the complement sample (dashed line) at FoM (or $\theta$) larger
  than the value on the x-axis.}
\end{figure}
The local p-value at this point is
$3.0\times10^{-4}$ and increases to $1.1\%$ after trial
correction. The scan in FoM and $\theta$ is shown in
Fig.~\ref{fig:2whsp_Pvalues}.

\begin{figure}
    \includegraphics[scale=1.0]{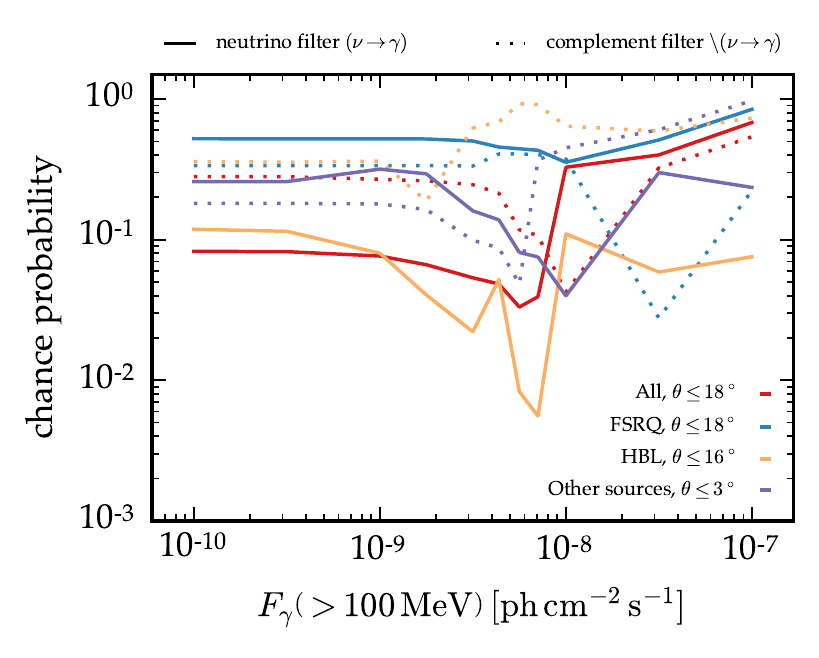}
    \includegraphics[scale=1.0]{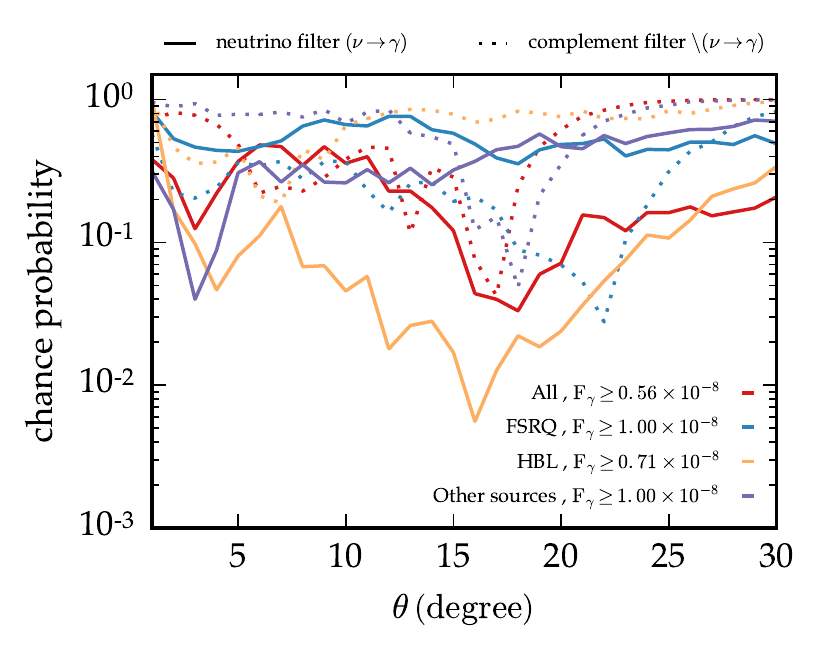}
    \caption{\label{fig:3lac_Pvalues} Result of the statistical test for
             partitions of the 3LAC catalogue over flux
             $F_\gamma\left(>100\:\mathrm{MeV}\right)$ (left) and CR angular
             distance $\theta$ (right). The probabilities are reported for the neutrino 
             filtered $\gamma$-ray emitters (solid line) and for the complement sample (dashed line) at $F_\gamma$ (or $\theta$) larger
  than the value on the x-axis.}
\end{figure}

The 3LAC catalogue~\cite{Ackermann:2015yfk} includes HBLs, FSRQ, and
other types of sources. As listed in Tab.~\ref{tab:results},
the most significant result is
again obtained for objects of the HBL type, consistent with the result of the 2FHL
catalogue. The scan in $F_\gamma$ and $\theta$ is shown in
Fig.~\ref{fig:3lac_Pvalues}. The most significant correlation is observed for
HBL objects with flux  $F_\gamma\left(>100\:\mathrm{MeV}\right)\geq0.71\times10^{-8}\:\mathrm{ph\,cm^{-2}\,s^{-1}}$
and $\theta \leq 17^\circ$. 131 CR events are observed
above an expectation of 93.4. The local p-value of this observation yields
$5.6\times10^{-3}$. After trial correction this increases to $11\%$. The
other types of objects in the catalogue (FSRQ and others) show a non significant
correlation.

\begin{figure}
    \includegraphics[scale=1.0]{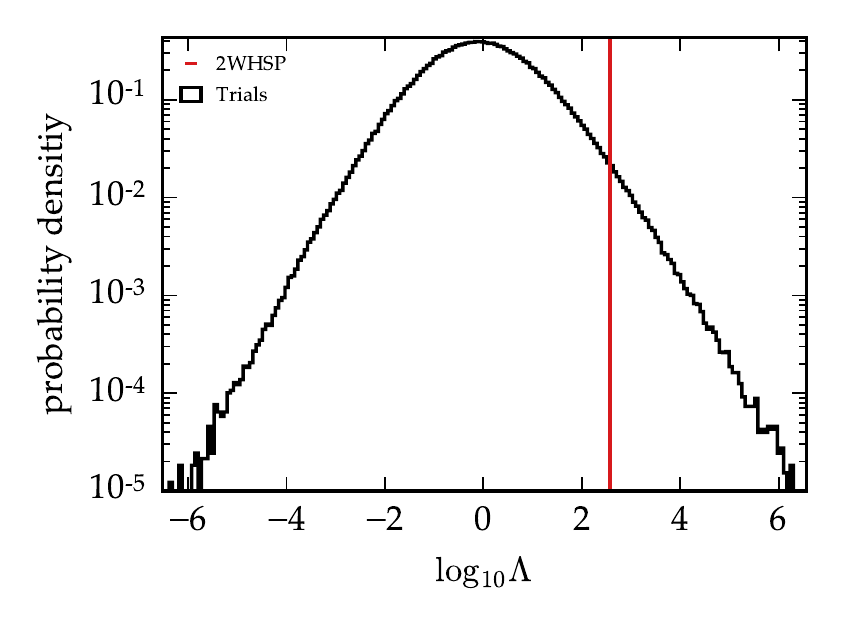}
    \includegraphics[scale=1.0]{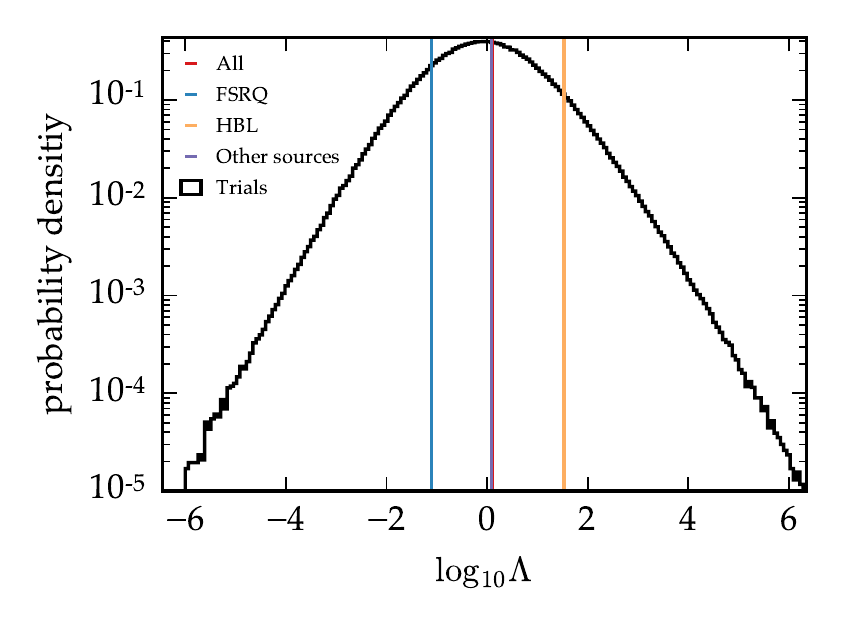}
    \caption{\label{fig:2whsp_3lac_logl} Result of the likelihood ratio test
             $\Lambda$ (Eq.~\ref{eq:llh_ratio})
             for the 2WHSP (left)
             and partitions of the 3LAC catalogue (right). Vertical lines
             indicate the outcome of the test using a catalogue with respect to
             the distribution of random trials (black histogram) as listed in
             Tab.~\ref{tab:results}.
             }
\end{figure}

The outcome of the likelihood ratio test $\Lambda$ (Eq.~(1))
is shown in Fig.~\ref{fig:2whsp_3lac_logl} for the 2WHSP (left) and 3LAC
catalogue (right). The test for the 2WHSP indicates a result in favour of neutrino filtered sources
with a p-value $1.2\%$.
For 3LAC, the result is $8.2\%$, again more significant for HBLs than for the
remaining subsets of the catalog.\\

\section{CRs in correlation with 2FHL objects.} The most significant result 
was seen for neutrinos and CRs correlating with 2FHL objects of the HBL type and
unclassified. Tab.~\ref{tab:counterparts} lists the clusters of neutrinos
correlating with 2FHL objects and CRs. Furthermore,
information about the events and objects is listed. Namely: the neutrino
energy and angular resolution, the redshift of the object (if known), the
position of the CR event, plus its energy and distance from the 2FHL object.

\clearpage
\begingroup
\onecolumn
\small
\begin{longtable}{ccc | llc |  cccccc}
\caption{IceCube neutrinos,  2FHL objects (type HBL and unclassified) and
         Auger/TA CRs selected in correspondence with the minimum p-value
         obtained ($\Fgamma\geq1.8\times10^{-11}\:\Funit$ and
         $\theta\leq10^{\circ}$).  We report for each neutrino the ID number,
         the energy and the median angular uncertainty
         (median($\theta)_{\textrm{IC}}$)  as quoted by IceCube; for the
         $\gamma$-ray objects the {\it Fermi} source name and common name, and
         if known, the redshift; for the CRs  the year of detection, the
         coordinates, the dataset, the energy and angular distance from the
         2FHL object.
\label{tab:counterparts}}\\
\multicolumn{3}{c}{High Energy Neutrino}&\multicolumn{3}{c}{$\gamma$-ray Source}&\multicolumn{6}{c}{ Ultra High Energy Cosmic Ray}  \\
\hline
ID&Energy&median($\theta)_{\textrm{IC}}$&2FHL name&Common name&z&Year&
RA&Dec
&Dataset&Energy&$\theta$ \\
&(TeV)&(deg)&&&&&&&&(EeV)&(deg)\\
\hline\endfirsthead
\caption{continued}\\
\multicolumn{3}{c}{High Energy Neutrino}&\multicolumn{3}{c}{$\gamma$-ray Source}&\multicolumn{6}{c}{ Ultra High Energy Cosmic Ray}  \\
\hline
ID&Energy&median($\theta)_{\textrm{IC}}$&2FHL name&Common name&z&Year&
RA&Dec
&Dataset&Energy&$\theta$ \\
&(TeV)&(deg)&&&&&&&&(EeV)&(deg)\\
\hline\endhead
9  & 63.2  & 16.5 & J0915.9+2931  & B2 0912+29             & 0.19 & 2013 & 138.6 & 26.1  & Auger & 62.1  & 3.4\\*
   &       &      &               &                        &      & 2008 & 140.0 & 28.7  & TA    & 59.2  & 1.0\\*
   &       &      &               &                        &      & 2010 & 129.0 & 29.1  & TA    & 60.5  & 8.5\\
   &       &      & J0910.4+3327  & Ton 1015               & 0.35 & 2013 & 138.6 & 26.1  & Auger & 62.1  & 7.4\\*
   &       &      &               &                        &      & 2008 & 140.0 & 28.7  & TA    & 59.2  & 5.1\\*
   &       &      &               &                        &      & 2010 & 145.0 & 40.7  & TA    & 92.2  & 9.1\\*
   &       &      &               &                        &      & 2010 & 137.0 & 41.5  & TA    & 68.9  & 8.1\\*
   &       &      &               &                        &      & 2010 & 129.0 & 29.1  & TA    & 60.5  & 8.3\\*
   &       &      & J1104.4+3812  & MKN 421                & 0.03 & 2011 & 163.7 & 28.9  & TA    & 62.3  & 9.5\\*
   &       &      &               &                        &      & 2011 & 157.0 & 38.8  & TA    & 72.9  & 7.3\\*
   &       &      &               &                        &      & 2012 & 160.0 & 35.6  & TA    & 57.4  & 5.7\\
   &       &      & J1015.0+4926  & 1ES 1011+496           & 0.20 & 2008 & 152.4 & 45.8  & TA    & 79.3  & 3.7\\*
   &       &      &               &                        &      & 2010 & 139.0 & 49.6  & TA    & 63.7  & 9.5\\*
   &       &      &               &                        &      & 2013 & 165.0 & 52.4  & TA    & 62.5  & 7.8\\
11 & 88.4  & 16.7 & J1027.0--1749 & 1RXS J102658.5--174905 & 0.65 & 2009 & 147.2 & -18.3 & Auger & 64.1  & 9.1\\*
   &       &      &               &                        &      & 2011 & 150.0 & -10.3 & Auger & 100.0 & 9.9\\*
   &       &      &               &                        &      & 2011 & 149.0 & -13.0 & Auger & 57.2  & 9.0\\*
   &       &      &               &                        &      & 2013 & 154.0 & -15.8 & Auger & 53.9  & 3.1\\
   &       &      & J0952.9--0841 & 1RXS J095303.4--084003 & --   & 2006 & 142.3 & -13.1 & Auger & 54.0  & 7.3\\*
   &       &      &               &                        &      & 2009 & 147.0 & -18.3 & Auger & 64.1  & 9.6\\*
   &       &      &               &                        &      & 2011 & 150.0 & -10.3 & Auger & 100.0 & 2.4\\*
   &       &      &               &                        &      & 2011 & 149.0 & -13.0 & Auger & 57.2  & 4.3\\*
   &       &      &               &                        &      & 2013 & 154.0 & -15.8 & Auger & 53.9  & 9.3\\
12 & 104   & 9.8  & J2009.4--4849 & PKS 2005--489          & 0.07 & 2006 & 305.6 & -46.3 & Auger & 60.0  & 3.3\\*
   &       &      &               &                        &      & 2007 & 315.0 & -53.8 & Auger & 72.7  & 9.4\\*
   &       &      &               &                        &      & 2008 & 306.0 & -55.1 & Auger & 55.1  & 6.8\\
   &       &      & J1959.6--4725 & SUMSS J195945--472519  & --   & 2006 & 305.6 & -46.3 & Auger & 60.1  & 4.0\\*
   &       &      &               &                        &      & 2008 & 306.0 & -55.1 & Auger & 55.1  & 8.7\\*
   &       &      &               &                        &      & 2013 & 308.0 & -39.5 & Auger & 67.3  & 9.9\\
   &       &      & J1936.9--4721 & PMN J1936--4719        & 0.26 & 2006 & 305.6 & -46.3 & Auger & 60.1  & 7.8\\*
   &       &      &               &                        &      & 2013 & 287.0 & -55.0 & Auger & 52.9  & 9.0\\
14 & 1041  & 13.2 & J1713.9--2027 & 1RXS J171405.2--202747 & --   & 2008 & 252.7 & -22.7 & Auger & 64.2  & 5.9\\
   &       &      & J1823.6-3454  & NVSS J182338--345412   & --   & 2013 & 284.5 & -37.6 & Auger & 54.4  & 7.4\\*
   &       &      &               &                        &      & 2009 & 286.0 & -37.8 & Auger & 61.0  & 8.8\\*
   &       &      &               &                        &      & 2009 & 276.0 & -33.4 & Auger & 65.8  & 1.5\\*
   &       &      &               &                        &      & 2011 & 284.0 & -28.6 & Auger & 80.9  & 9.2\\
   &       &      & J1829.0--2417 & 1RXS J182853.8--241746 & --   & 2010 & 284.7 & -28.2 & Auger & 65.2  & 7.6\\*
   &       &      &               &                        &      & 2009 & 276.0 & -33.4 & Auger & 65.8  & 9.1\\*
   &       &      &               &                        &      & 2011 & 284.0 & -28.6 & Auger & 80.9  & 7.1\\
   &       &      & J1741.2--4021 &                        & --   & 2010 & 258.1 & -44.9 & Auger & 72.9  & 7.0\\*
   &       &      &               &                        &      & 2012 & 260.0 & -32.7 & Auger & 61.8  & 8.9\\
17 & 200   & 11.6 & J1555.7+1111  & PG 1553+113            & 0.44 & 2007 & 245.8 & 8.5   & Auger & 54.9  & 7.3\\*
   &       &      &               &                        &      & 2011 & 239.0 & 3.9   & Auger & 60.3  & 7.3\\
19 & 71.5  & 9.7  & J0543.9--5533 & 1RXS J054357.3--553206 & 0.27 & 2010 & 80.2  & -64.1 & Auger & 54.3  & 9.0\\*
   &       &      &               &                        &      & 2013 & 92.1  & -64.1 & Auger & 65.4  & 9.1\\*
   &       &      &               &                        &      & 2013 & 91.4  & -60.6 & Auger & 72.5  & 5.8\\
20 & 1141  & 10.7 & J0352.7--6831 & PKS 0352--686          & 0.08 & 2012 & 37.0  & -75.8 & Auger & 58.7  & 9.6\\*
   &       &      &               &                        &      & 2014 & 45.2  & -65.8 & Auger & 63.6  & 5.7\\*
   &       &      &               &                        &      & 2010 & 80.2  & -64.1 & Auger & 54.3  & 9.8\\*
   &       &      &               &                        &      & 2013 & 56.6  & -67.8 & Auger & 70.7  & 0.9\\*
   &       &      &               &                        &      & 2013 & 64.7  & -70.1 & Auger & 68.8  & 2.8\\*
   &       &      &               &                        &      & 2014 & 72.8  & -73.5 & Auger & 60.0  & 6.8\\
22 & 220   & 12.1 & J1958.3--3011 & 1RXS J195815.6--30111  & 0.12 & 2011 & 295.1 & -27.6 & Auger & 54.8  & 4.7\\*
   &       &      &               &                        &      & 2008 & 304.0 & -26.2 & Auger & 52.6  & 5.8\\*
   &       &      &               &                        &      & 2011 & 305.0 & -34.5 & Auger & 67.4  & 6.6\\
   &       &      & J1917.7--1921 & 1H1914--194            & 0.14 & 2010 & 284.7 & -28.2 & Auger & 65.2  & 9.8\\*
   &       &      &               &                        &      & 2011 & 295.0 & -27.6 & Auger & 54.8  & 9.8\\*
   &       &      &               &                        &      & 2009 & 294.0 & -20.5 & Auger & 59.5  & 4.9\\
   &       &      & J1921.9--1607 & PMNJ1921--1607         & --   & 2009 & 294.5 & -20.5 & Auger & 59.5  & 5.8\\
26 & 210   & 11.8 & J0905.7+1359  & MG1 J090534+1358       & 1.07 & 2011 & 132.8 & 12.9  & Auger & 55.9  & 3.7\\*
   &       &      &               &                        &      & 2007 & 137.0 & 6.2   & Auger & 53.6  & 7.9\\*
   &       &      &               &                        &      & 2009 & 129.0 & 15.2  & Auger & 52.2  & 6.9\\
   &       &      & J0915.9+2931  & B2 0912+29             & 0.19 & 2013 & 138.6 & 26.1  & Auger & 62.1  & 3.4\\*
   &       &      &               &                        &      & 2008 & 140.0 & 28.7  & TA    & 59.2  & 1.0\\*
   &       &      &               &                        &      & 2010 & 129.0 & 29.1  & TA    & 60.5  & 8.5\\
27 & 60.2  & 6.6  & J0816.3--1311 & PMN J0816--1311        & --   & 2010 & 131.9 & -15.5 & Auger & 76.1  & 7.9\\*
   &       &      &               &                        &      & 2013 & 123.0 & -6.2  & Auger & 85.3  & 7.0\\
33 & 385   & 13.5 & J1933.3+0725  & 1RXS J193320.3+072616  & --   & 2008 & 287.7 & 1.5   & Auger & 118.0 & 8.2\\*
   &       &      &               &                        &      & 2013 & 299.0 & 8.7   & Auger & 54.6  & 5.4\\*
   &       &      &               &                        &      & 2011 & 288.0 & 0.3   & TA    & 136.0 & 8.8\\
   &       &      & J1931.1+0937  & RX J1931.1+0937        & --   & 2008 & 287.7 & 1.5   & Auger & 118.3 & 9.6\\*
   &       &      &               &                        &      & 2013 & 299.0 & 8.7   & Auger & 54.6  & 5.9\\
   &       &      & J1942.8+1033  & 1RXS J194246.3+103339  & --   & 2006 & 299.0 & 19.4  & Auger & 82.0  & 9.4\\*
   &       &      &               &                        &      & 2013 & 299.0 & 8.7   & Auger & 54.6  & 3.5\\
35 & 2004  & 15.9 & J1328.6--4728 & 1WGA J1328.6--4727     & --   & 2005 & 199.1 & -48.5 & Auger & 52.1  & 2.3\\*
   &       &      &               &                        &      & 2006 & 201.0 & -55.3 & Auger & 69.5  & 7.8\\*
   &       &      &               &                        &      & 2006 & 201.0 & -45.3 & Auger & 59.5  & 2.4\\*
   &       &      &               &                        &      & 2007 & 200.0 & -43.4 & Auger & 60.0  & 4.3\\*
   &       &      &               &                        &      & 2008 & 202.0 & -54.9 & Auger & 53.4  & 7.4\\
   &       &      & J1304.5--4353 & 1RXS 130421.2--435308  & --   & 2004 & 199.7 & -34.8 & Auger & 84.7  & 9.4\\*
   &       &      &               &                        &      & 2005 & 199.0 & -48.5 & Auger & 52.1  & 5.1\\*
   &       &      &               &                        &      & 2006 & 201.0 & -45.3 & Auger & 59.5  & 3.7\\*
   &       &      &               &                        &      & 2007 & 200.0 & -43.4 & Auger & 60.0  & 3.0\\*
   &       &      &               &                        &      & 2007 & 193.0 & -35.3 & Auger & 60.7  & 9.0\\*
   &       &      &               &                        &      & 2009 & 194.0 & -36.4 & Auger & 72.5  & 7.7\\
   &       &      & J1307.6--4259 & 1RXS 130737.8--425940  & --   & 2004 & 199.7 & -34.8 & Auger & 84.7  & 8.4\\*
   &       &      &               &                        &      & 2005 & 199.0 & -48.5 & Auger & 52.1  & 5.8\\*
   &       &      &               &                        &      & 2006 & 201.0 & -45.3 & Auger & 59.5  & 3.7\\*
   &       &      &               &                        &      & 2007 & 200.0 & -43.4 & Auger & 60.0  & 2.4\\*
   &       &      &               &                        &      & 2007 & 193.0 & -35.3 & Auger & 60.7  & 8.4\\*
   &       &      &               &                        &      & 2009 & 194.0 & -36.4 & Auger & 72.5  & 7.0\\*
   &       &      &               &                        &      & 2013 & 201.0 & -34.6 & Auger & 62.7  & 9.0\\
   &       &      & J1353.5--6640 & 1RXS J135341.1--664002 & --   & 2007 & 195.5 & -63.4 & Auger & 61.9  & 6.3\\*
   &       &      &               &                        &      & 2008 & 196.0 & -69.7 & Auger & 71.1  & 5.5\\*
   &       &      &               &                        &      & 2013 & 199.0 & -63.9 & Auger & 53.2  & 4.9\\*
   &       &      &               &                        &      & 2004 & 208.0 & -60.1 & Auger & 58.6  & 6.6\\*
   &       &      &               &                        &      & 2008 & 187.0 & -63.5 & Auger & 65.3  & 9.3\\*
   &       &      &               &                        &      & 2010 & 216.0 & -66.5 & Auger & 60.3  & 3.1\\*
   &       &      &               &                        &      & 2010 & 219.0 & -70.8 & Auger & 89.1  & 5.6\\
   &       &      & J1507.4--6213 &                        & --   & 2013 & 240.3 & -68.9 & Auger & 61.5  & 8.7\\*
   &       &      &               &                        &      & 2004 & 208.0 & -60.1 & Auger & 58.6  & 9.3\\*
   &       &      &               &                        &      & 2007 & 220.0 & -53.9 & Auger & 61.5  & 9.2\\*
   &       &      &               &                        &      & 2010 & 216.0 & -66.5 & Auger & 60.3  & 6.3\\*
   &       &      &               &                        &      & 2010 & 219.0 & -70.8 & Auger & 89.1  & 9.2\\*
   &       &      &               &                        &      & 2010 & 232.0 & -56.6 & Auger & 54.9  & 6.2\\
39 & 101.3 & 14.2 & J0649.6--3139 & 1RXSJ064933.8--31391   & --   & 2007 & 105.9 & -22.8 & Auger & 60.8  & 9.3\\
   &       &      & J0622.4--2604 & PMNJ0622--2605         & --   & 2007 & 105.9 & -22.8 & Auger & 60.8  & 9.9\\
   &       &      & J0631.0--2406 & 1RXSJ063059.7--240636  & --   & 2007 & 105.9 & -22.8 & Auger & 60.8  & 7.6\\
   &       &      & J0639.9--1252 &                        & --   & --   & -     & -     & --    & -     & -\\
41 & 87.6  & 11.1 & J0416.9+0105  & 1ES 0414+009           & 0.29 & 2008 & 67.7  & 4.0   & Auger & 52.0  & 4.5\\*
   &       &      &               &                        &      & 2012 & 56.4  & -3.2  & Auger & 53.3  & 9.0\\
46 & 158   & 7.6  & J1027.0--1749 & 1RXS J102658.5--174905 & 0.65 & 2009 & 147.2 & -18.2 & Auger & 64.1  & 9.1\\*
   &       &      &               &                        &      & 2011 & 150.0 & -10.3 & Auger & 100.0 & 9.9\\*
   &       &      &               &                        &      & 2011 & 149.0 & -13.0 & Auger & 57.2  & 9.0\\*
   &       &      &               &                        &      & 2013 & 154.0 & -15.8 & Auger & 53.9  & 3.1\\*
48 & 104.7 & 8.1  & J1440.7--3847 & 1RXS J144037.4--38465  & --   & 2004 & 224.7 & -44.0 & Auger & 58.2  & 6.2\\*
   &       &      &               &                        &      & 2008 & 221.0 & -42.8 & Auger & 73.1  & 4.0\\*
   &       &      &               &                        &      & 2011 & 219.0 & -41.9 & Auger & 58.8  & 3.2\\
51 & 66.2  & 6.5  & J0540.5+5822  & GB6 J0540+5823         & --   & 2009 & 99.2  & 62.7  & TA    & 80.7  & 8.1\\*
   &       &      &               &                        &      & 2010 & 78.8  & 61.4  & TA    & 61.2  & 4.4\\*
   &       &      &               &                        &      & 2011 & 82.5  & 57.7  & TA    & 74.7  & 1.6\\
\hline
\end{longtable}
\endgroup

\bibliographystyle{mnras}
\bibliography{neutrino}
\label{lastpage}

\bsp	

\end{document}